\documentclass[a4paper]{revtex4}
\usepackage{epsfig}
\usepackage{amsmath}
 \newtheorem{Proposition}{Proposition}
 
 \newtheorem{Theorem}{Theorem}
 
 \newtheorem{Definition}{Definition}
\usepackage{amssymb}

\begin{document}

\def\myint{\int \!\! d^{ {\scriptscriptstyle D} } \boldsymbol{x} \; }

\title{Existence of dark solitons in a class of stationary nonlinear Schr\"odinger equations with periodically modulated nonlinearity and periodic asymptotic}

\author{J. Belmonte-Beitia$^1$ and J. Cuevas$^2$}

\affiliation{
$^1$ Departamento de Matem\'aticas, E. T. S. de Ingenieros Industriales and Instituto de Matem\'atica Aplicada a la Ciencia y la Ingenier\'{\i}a (IMACI), \\ Universidad de Castilla-La Mancha 13071 Ciudad Real, Spain.\\
$^2$ Grupo de F\'{\i}sica No Lineal. Universidad de Sevilla\\ Departamento de F\'{\i}sica Aplicada I. Escuela Polit\'ecnica Superior. \\
C/ Virgen de \'Africa, 7. 41011 Sevilla, Spain.}

\date{\today}

\begin{abstract}
\noindent In this paper, we give a proof of the existence of stationary dark soliton solutions or heteroclinic orbits of nonlinear equations of
Schr\"odinger type with periodic inhomogeneous nonlinearity. The result is illustrated with examples of dark solitons for cubic and photorefractive nonlinearities.
\end{abstract}


\maketitle

\section{Introduction}

Nonlinear Schr\"odinger (NLS) equations have a great number of applications in different fields of modern physics, from laser wavepackets propagating in nonlinear materials to matter waves in Bose-Einstein condensates (BEC), gravitational models for quantum mechanics, plasma physics or wave propagation in biological and geological systems  \cite{AA,Sulembook,PG}.

From the physical point of view two of the most relevant forms of the nonlinear equations of Schr\"odinger type are the stationary NLS equation with cubic nonlinearity (namely cubic NLS),
\begin{equation}\label{eq1}
-u''+\lambda u+au^3=0
\end{equation}
and the stationary NLS equation with cubic-quintic nonlinearity (namely cubic-quintic NLS),
\begin{equation}\label{eq2}
-u''+\lambda u+au^3+bu^5=0,
\end{equation}
where $a,b\in\mathbb{R}$. In the case when $a<0$, for the first model (respectively, $b<0$ for the second model), and with boundary conditions
\begin{equation}
\lim_{|x|\rightarrow\infty}u(x)=0,\quad \lim_{|x|\rightarrow\infty}u'(x)=0,
\end{equation}
the solutions of Eq. (\ref{eq1}) are called solitary waves or bright solitons. When $a>0$, (respectively, $b>0$) nontrivial dynamics requires that the solution does not vanish at infinity. In this case the so-called dark solitons or heteroclinic orbits appear.

In the last years there has been an increasing interest in the one-dimensional nonlinear Schr\"odinger equation with a spatially modulated nonlinearity:
 \begin{equation}
 \label{NLSEinh}
-u''+\lambda u+a(x)u^3=0
 \end{equation}
 with $x\in \mathbb{R}$ and $a(x)$ describing the spatial modulation of the nonlinearity. This equation arises in different physical contexts such as nonlinear optics and Bose-Einstein condensation, being in the later field which has gained more interest in the last year. This fact has been triggered by the possibility of using the Feschbach resonance management techniques to modify spatially the collisional interactions between atoms
 \cite{Victor1,Garnier,Panos2}. Different aspects of the dynamics of solitons in these frameworks have been considered such as the emission of solitons \cite{Victor1} and the soliton propagation when the space modulation of the nonlinearity is either a random \cite{Garnier}, periodic \cite{Boris}, linear \cite{Panos2} or localized function \cite{Primatarowa}.

In recent years, the study of existence of bright solitary waves has gained a lot of interest from the mathematical point of view (see e.g. \cite{Ambrosetti,Lions,Strauss,ptorres}). The existence of dark solitons has been previously considered in non-periodic potentials (see \cite{Panos} and references therein). To the best of our knowledge, the only proof of existence in NLS models with periodic potentials (optical lattices) has been presented for the cubic-quintic nonlinearity in \cite{Pedro}.

Many NLS models, including Eq. (\ref{NLSEinh}), can be considered as specific cases of a more general equation with a periodic nonlinearity coefficient
\begin{equation}\label{estacionario}
    -u_{xx}+\lambda u+g(x)f(u)u=0,
\end{equation}
with $g: \mathbb{R}\rightarrow\mathbb{R}$ being $T$-periodic, satisfying the following properties:
\begin{subequations}
 \begin{eqnarray}
 \label{conditions1} 0<g_{\min}\leqslant g(x)\leqslant g_{\max}
\\
\label{conditions2}  g(x) = g(-x).
\end{eqnarray}
\end{subequations}
and $f:\mathbb{R}\rightarrow\mathbb{R}$ satisfying
\begin{equation}\label{conditions3}
f \in C^{1}(\mathbb{R}),\quad f(-u)=f(u)\geq 0, u\in\mathbb{R},\quad\text{and}\quad f'(u)>0, u\in\mathbb{R}_{+}.\\
\end{equation}

Usually a dark soliton is defined as a solution of Eq. (\ref{estacionario}) verifying the following
asymptotic boundary conditions
\begin{equation}
\label{boundary} \frac{u(x)}{u_{\pm}(x)}\rightarrow 1,
\qquad x\rightarrow{\pm\infty},
\end{equation}
where the functions $u_{\pm}(x)$ are sign definite, $T$-periodic, real solutions of Eq. (\ref{estacionario}). By symmetry of Eq. (\ref{estacionario}), $u_{-}(x)=-u_{+}(x)$, which allows to consider odd solutions for $u(x)$ and to reduce complexity of the problem. This kind of solutions are also denoted as asymptotically oscillatory kinks or black solitons and possess a topological charge. Notice that these solutions are different to bubbles, for which $\lim_{x\rightarrow+\infty}u(x)= \lim_{x\rightarrow-\infty}u(x)$ \cite{B1,B2,B3}.

The aim of this paper is, therefore, to extend earlier studies to the model given by (\ref{estacionario}), which, to the best of our knowledge, has not previously been considered. From the mathematical point of view, the strategy of proof consists of several techniques from the classical  theory of ODE's (upper and lower solutions) and planar homeomorphisms (topological degree and free homeomorphisms) combined in a novel way.

The rest of the paper is organized as follows. In Section 2 the preliminaries of the paper are introduced and a proof of the existence of periodic solutions for Eq. (\ref{estacionario}) is given. Section 3 contains the main result about the existence of heteroclinic orbits or dark solitons. In section 4 the theorems of Section 3 are applied to two typical examples: the cubic and photorefractive nonlinearities. We conclude in Section 5 with a summary of the main findings of the paper.

\section{Existence of periodic solutions}

We start by analyzing the range of values of $\lambda$ for which the existence of nontrivial solutions of Eq. (\ref{estacionario}) can be found.

\begin{Theorem}
If $\lambda\geqslant 0$, the only bounded solution of Eq.
(\ref{estacionario}) is the trivial one, $u=0$.
\end{Theorem}
\textbf{Proof:}
Let $u$ be a nontrivial solution of Eq. (\ref{estacionario}).
Multiplying (\ref{estacionario}) by $u$ and integrating over the whole space, one gets
\begin{equation}
\int \left((u_{x})^2+\lambda u^2+g(x)f(u)u^2\right) dx=0
\end{equation}
Using (\ref{conditions1}), (\ref{conditions2}) and, by hypothesis, the fact that $\lambda\geqslant0$, the only bounded solution is $u=0$.
\rule{4pt}{5pt}

Therefore, throughout  this paper, we will take $\lambda<0$. As
$g_{\text{min}}\leqslant g(x)\leqslant g_{\text{max}}$, let us consider two
auxiliary autonomous equations:
\begin{eqnarray}
\label{auxiliares1}-u^{(1)}_{xx}+\lambda u^{(1)}+g_{\min}f(u^{(1)})u^{(1)}=0 \\
\label{auxiliares2}-u^{(2)}_{xx}+\lambda u^{(2)}+g_{\max}f(u^{(2)})u^{(2)}=0
\end{eqnarray}
These equations possess two nontrivial equilibria
\begin{eqnarray}
\label{equilibria1} \xi^{(1)}=f^{-1}\left(-\frac{\lambda}{g_{\min}}\right),\\
\label{equilibria2} \xi^{(2)}=f^{-1}\left(-\frac{\lambda}{g_{\max}}\right).
\end{eqnarray}
We need to classify these points. Linearizing Eqs. (\ref{auxiliares1}) and (\ref{auxiliares2}) around of equilibria points (\ref{equilibria1}) and (\ref{equilibria2}), one obtains the following eigenvalues
\begin{eqnarray}
\mu^{(1)}&=&\pm\sqrt{g_{\text{min}}f'\left(f^{-1}\left(\frac{-\lambda}{g_{\min}}\right)\right)f^{-1}\left(\frac{-\lambda}{g_{\min}}\right)},\\
\mu^{(2)}&=&\pm\sqrt{g_{\text{max}}f'\left(f^{-1}\left(\frac{-\lambda}{g_{\max}}\right)\right)f^{-1}\left(\frac{-\lambda}{g_{\max}}\right)}
\end{eqnarray}
As $f$ is a continuous and increasing function, there exists $f^{-1}$. Moreover $f^{-1}$ is also an increasing function. Thus, the points (\ref{equilibria1}), (\ref{equilibria2}) are saddle points. Moreover, due to the evenness of the function $f$, there exist two extra equilibrium points more, $\xi^{(3)}=-\xi^{(1)}$ and $\xi^{(4)}=-\xi^{(2)}$. We would also like to note that $\xi^{(1)}>\xi^{(2)}$, since that $f$ is increasing on the positive semi-axis.

Before continuing, let us recall a useful definition from the theory of ODE (see, for example \cite{Coster2}). Let the following second order differential equation be
\begin{equation}
\label{ODE}
u_{xx}=h(x,u),
\end{equation}
where $h$ is a continuous function with respect to both arguments and $T$-periodic in $x$.

\begin{Definition}

(i) We say that $\bar{u}: [a,+\infty)\rightarrow\mathbb{R}$ is a lower solution of (\ref{ODE}) if
\begin{equation}
\bar{u}_{xx}> h(x,\bar{u})
\end{equation}
for all $x>a$.

(ii) Similarly, $\underline{u}: [a,+\infty)\rightarrow\mathbb{R}$ is an upper solution of (\ref{ODE}) provided that
\begin{equation}
\underline{u}_{xx}< h(x,\underline{u})
\end{equation}
for all $x>a$.
\end{Definition}

We shall now prove the existence of a $T$-periodic and an
unstable solution between both points $\xi^{(1)}$ and $\xi^{(2)}$.

\begin{Proposition}\label{pr1}
The points $\xi^{(1)}$ and $\xi^{(2)}$, which were previously calculated, are respectively constant upper and lower solutions of Eq
(\ref{estacionario}). Moreover there exists an unstable periodic
solution situated between them.
\end{Proposition}
\emph{Proof}
By using Eq. (\ref{auxiliares1}), we obtain:
\begin{equation}
-\xi^{(1)}_{xx}+\lambda\xi^{(1)}+g(x)f(\xi^{(1)})\xi^{(1)}>\lambda\xi^{(1)}+g_{min}f(\xi^{(1)})\xi^{(1)} =0
\end{equation}
and, similarly, for the Eq. (\ref{auxiliares2}):
\begin{equation}
-\xi^{(2)}_{xx}+\lambda\xi^{(2)}+g(x)f(\xi^{(2)})\xi^{2}<\lambda\xi^{(2)}+g_{max}f(\xi^{(2)})\xi^{(2)} =0
\end{equation}
Thus, by the latter definition, $\xi^{(1)}$ and $\xi^{(2)}$ are a couple of well-ordered
upper and lower solutions, respectively. Therefore, there exists
 a $T$-periodic solution between them \cite{Coster2}. Such solution is unstable because the associated Brouwer index to the Poincare map is
$-1$ (see,
for example \cite{Ortega}).
\rule{5pt}{4pt}

We have therefore a positive and $T$-periodic solution of Eq. (\ref{estacionario}), $u_{+}(x)$, satisfying $\xi^{(2)}\leqslant u_{+}(x)\leqslant\xi^{(1)}$. Due to the symmetry of the equation we also have a negative solution $u_{-}(x)=-u_{+}(x)$.

\section{Existence of a dark soliton.}

In this section, we prove the existence of a heteroclinic orbit connecting the periodic solutions $u_{-}$ and $u_{+}$.
This heteroclinic orbit can also be called a dark soliton.

The following theorem, proved in \cite{Pedro} with the use of ideas from \cite{Pedro1}, is the key one for this work.
\begin{Theorem}
\label{to2}
Let $w,v: [a,+\infty)\rightarrow\mathbb{R}$ be a bounded functions verifying the following conditions
\begin{enumerate}

\item \label{c1} $w(x)<v(x), \quad\forall x>a$

\item \label{c2} $w_{xx}(x)>h(x,w)$ and $v_{xx}(x)<h(x,v),\quad\forall x>a$.

Then there exists a solution $u(x)$ of (\ref{ODE}) such that\begin{equation}
w(x)<u(x)<v(x)
\end{equation}
If moreover, there exists $x$ such that
\item \label{c3} $\underset{\substack{x\in[0,T] \\ y\in[\inf_{x\geqslant a}w(x), \sup_{x\geqslant a}v(x)]}}{\min} \frac{\partial h(x,y)}{\partial y}>0$,

\end{enumerate}

then there exists an $T$-periodic solution $\rho(x)$ such that
\begin{equation}
\lim_{x\rightarrow+\infty}(|u(x)-\rho(x)|+|u_{x}(x)-\rho_{x},(x)|)=0
\end{equation}
Moreover, $\rho(x)$ is the unique T-periodic solution in the interval $[\inf_{x\geqslant x_{0}}w(x),\sup_{x\geqslant x_{0}}v(x)]$.
\end{Theorem}

Let us apply this theorem to our model. We have that
\begin{equation}
h(x,u)=\lambda u(x)+g(x)f(u(x))u.
\end{equation}
Since $g(x)$ is symmetric, one can consider $x\geqslant0$ and
extend the obtained solution $u(x)$ as an odd function for
$x<0$. The solutions of Eqs (\ref{auxiliares1}) and
(\ref{auxiliares2}), $u^{(1)}$ and $u^{(2)}$, which are
heteroclinic orbits joining $-\xi^{(1)}$ with $\xi^{(1)}$ and
$-\xi^{(2)}$ with $\xi^{(2)}$, respectively, satisfy the
conditions (\ref{c1}) and (\ref{c2}) of Theorem \ref{to2}, with
$v(x)=u^{(1)}(x)$ and $w(x)=u^{(2)}(x)$. We thus have a
bounded solution $u(x)$ of Eq. (\ref{estacionario}) such that
\begin{equation}
u^{(2)}(x)<u(x)<u^{(1)}(x)
\end{equation}
Hence, in order to prove that there exists a solution $u(x)$ of (\ref{estacionario}) converging to $u_{+}(x)$ and $u_{-}(x)$, which were obtained in Proposition \ref{pr1}, as $x\rightarrow\pm\infty$, one has to verify the condition (\ref{c3}) of Theorem \ref{to2}.

As $a$ can be taken arbitrarily large, condition (\ref{c3}) is equivalent to
\begin{equation}
\label{eqmin} \underset{\substack{x\in[0,T] \\
u\in[\xi^{(2)},\xi^{(1)}]}}{\min}[\lambda+g(x) f(u)+g(x)f'(u)u]>0
\end{equation}

Here, we must distinguish four different cases:

\begin{description}

\item{(i)} $f'(u)$ is an increasing function, for $u>0$. Then (\ref{eqmin}) is now equivalent to
\begin{equation}
\lambda+g_{\min}f(\xi^{(2)})+g_{\min}f'(\xi^{(2)})\xi^{(2)}>0
\end{equation}
Taking into account Eqs. (\ref{equilibria1}) and (\ref{equilibria2}), a straightforward analysis of the last inequality gives the following estimate for $\lambda$
\begin{equation}\label{condicion1}
\lambda-\lambda\frac{g_{\min}}{g_{\max}}+g_{\min}f'\left(f^{-1}(-\frac{\lambda}{g_{\max}})\right)f^{-1}(-\frac{\lambda}{g_{\max}})>0
\end{equation}
As $f$ is an increasing function, condition (\ref{condicion1}) can be satisfied for certain values of $g_{\min}$, $g_{\max}$, $\lambda$ and, therefore, it is a sufficient condition for the existence of dark solitons.

\item{(ii)} $f'(u)$ is a decreasing function, for $u>0$. Then, constraint (\ref{eqmin}) is reduced to
\begin{equation}
\lambda+g_{\min}f(\xi^{(2)})+g_{\min}f'(\xi^{(1)})\xi^{(2)}>0
\end{equation}
and subsequently to the following inequality
\begin{equation}\label{condicion2}
\lambda-\lambda\frac{g_{\min}}{g_{\max}}+g_{\min}f'\left(f^{-1}(-\frac{\lambda}{g_{\min}})\right)f^{-1}(-\frac{\lambda}{g_{\max}})>0
\end{equation}
Again, condition (\ref{condicion2}) can be satisfied, depending on the values of $g_{\min}$, $g_{\max}$ and $\lambda$.

\item{(iii)} $f'(u)$ has, at least, a minimal point in $u_{0}\in[\xi^{(2)},\xi^{(1)}]$. For this case, the constraint (\ref{eqmin}) is reduced to
\begin{equation}
\lambda+g_{\min}f(\xi^{(2)})+g_{\min}f'(u_{0})\xi^{(2)}>0
\end{equation}
and hence
\begin{equation}\label{condicion3}
\lambda-\lambda\frac{g_{\min}}{g_{\max}}+g_{\min}f'(u_{0})f^{-1}(-\frac{\lambda}{g_{\max}})>0
\end{equation}
Thus (\ref{condicion3}) is a sufficient condition for the existence of dark solitons.

\item{(iv)} $f'(u)$ has a maximal point in $u_{1}\in[\xi^{(2)},\xi^{(1)}]$. Thus, (\ref{eqmin}) is reduced to
\begin{equation}\label{eq-1}
\lambda+g_{\min}f(\xi^{(2)})+g_{\min}f'(\xi^{(2)})\xi^{(2)}>0
\end{equation}
if we supposse that $f'(\xi^{(2)})<f'(\xi^{(1)})$. Otherwise, we hold
\begin{equation}\label{eq-2}
\lambda+g_{\min}f(\xi^{(2)})+g_{\min}f'(\xi^{(1)})\xi^{(2)}>0
\end{equation}
Thus, Eq. (\ref{eq-1}) gives the following estimate for $\lambda$
\begin{equation}
\lambda-\lambda\frac{g_{\min}}{g_{\max}}+g_{\min}f'\left(f^{-1}(-\frac{\lambda}{g_{\max}})\right)f^{-1}(-\frac{\lambda}{g_{\max}})>0
\end{equation}
and, alternatively, Eq. (\ref{eq-2}) gives the following estimate for $\lambda$
\begin{equation}
\lambda-\lambda\frac{g_{\min}}{g_{\max}}+g_{\min}f'\left(f^{-1}(-\frac{\lambda}{g_{\min}})\right)f^{-1}(-\frac{\lambda}{g_{\max}})>0
\end{equation}
which are Eqs. (\ref{condicion1}) and (\ref{condicion2}).
\end{description}

\section{Application to the existence of dark solitons}

In this section we check the consequences of our main result in the problem of the existence of dark solitons to the stationary nonlinear Schr\"odinger equation with periodic nonlinearity, given by Eq. (\ref{estacionario}).

We distinguish hereafter two cases of great physics relevance: the NLS with either cubic or photorefractive nonlinearity.

\subsection{Dark solitons in the cubic NLS equation}

The cubic nonlinearity, $f(u)=u^2$, is by far the most common in several fields of physics, as nonlinear optics \cite{Kivshar0} (where it is also dubbed as Kerr nonlinearity) and Bose-Einstein condensation \cite{Ricardo}. Dark solitons arise in self-defocusing optical Kerr media \cite{Kivshar2} or repulsive Bose-Einstein condensates \cite{Dimitri}.

Thus, if $f(u)=u^2$, Eq. (\ref{estacionario}) can be cast as
\begin{equation}\label{estacionario2}
-u_{xx}+\lambda u+g(x)u^3=0
\end{equation}
Therefore, we are in the case $(i)$ of the previous section. Thus, condition (\ref{c3}) is equivalent to
\begin{equation}
 \underset{\substack{x\in[0,T] \\
u\in[\xi^{(2)},\xi^{(1)}]}}{\min}[2\lambda+6g(x) u^{2}]>0
\end{equation}
This last inequality is equivalent to $2\lambda+6g_{\min}(\xi^{(2)})^{2}>0$. Using Eq. (\ref{equilibria2}) and the fact that $\lambda<0$, we find a
connection between $g_{\min}$ and $g_{\max}$:
\begin{equation}
\label{condicionfinal} g_{\min}>\frac{g_{\max}}{3}.
\end{equation}
Consequently, if (\ref{condicionfinal}) is verified, dark solitons will exist in the cubic nonlinear Sch\"odinger equation with periodic nonlinearity.

We consider below an example of dark (black) soliton of the Eq. (\ref{estacionario2}), in order to check the concepts introduced in Section 3. This example corresponds to the periodic nonlinearity $g(x)$ given by
\begin{equation}
    g(x)=\frac{g_{0}}{(1+\alpha\cos(\omega x))^{3}}
\end{equation}
with $\omega=2\sqrt{|\lambda|}$, $g_{0}$ and $\alpha<1$ being positive constants. In order to satisfy the relation (\ref{condicionfinal}), $\alpha$ must fulfill the constrain $\alpha<(3^{1/3}-1)/(3^{1/3}+1)$.
For this case, the solutions of Eqs. (\ref{auxiliares1}) and (\ref{auxiliares2}), for $f(u)=u^2$ can be found analitically, being
\begin{equation}
u^{(1)}=\sqrt{\frac{-\lambda}{g_{\min}}}\tanh\left(\sqrt{\frac{-\lambda}{2}}x\right),\quad u^{(2)}=\sqrt{\frac{-\lambda}{g_{\max}}}\tanh\left(\sqrt{\frac{-\lambda}{2}}x\right)
\end{equation}

\begin{figure}[t]
    \includegraphics[width=12cm]{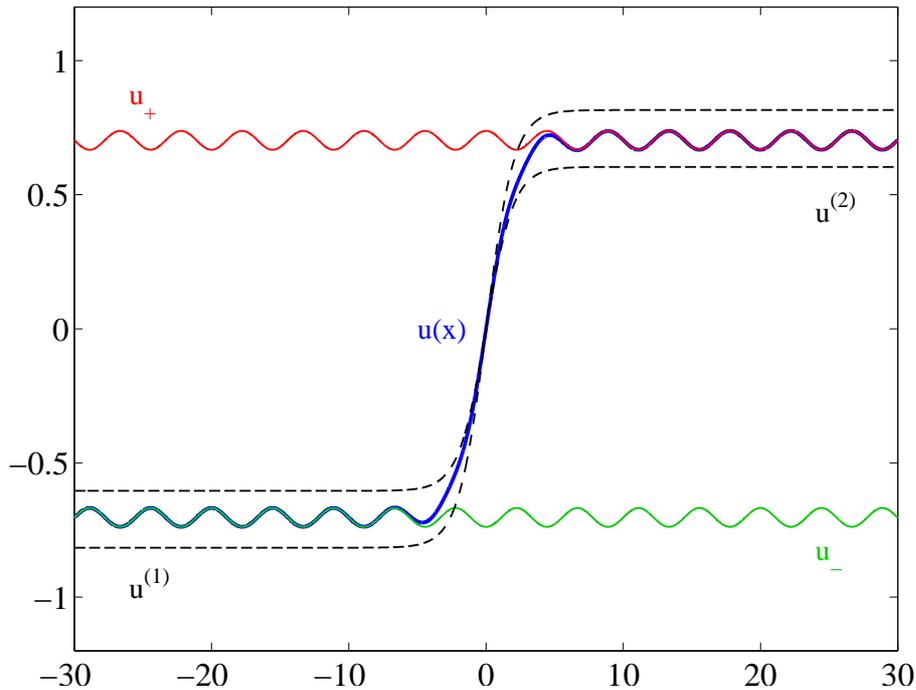}
\caption{[Color Online]. The dark soliton $u(x)$ (solid blue line), the hyperbolic periodic solutions $u_{\pm}(x)$ (dashed red and green line), and the heteroclinic orbits $u^{(1)}$ and $u^{(2)}$ (dashed black line) connecting $\pm\xi^{(1)}$ and $\pm\xi^{(2)}$ respectively, for the parameters $\lambda=-0.5$, $g_{0}=1$, $\alpha=0.1$ of the cubic NLS equation. \label{darksoliton}}
\end{figure}

Following \cite{Nuestro,Nuestro2}, the solution of Eq. (\ref{estacionario2}) with the boundary conditions (\ref{boundary2}) is
 \begin{equation}\label{brito}
 u(x)  =
 \frac{\omega}{2}
 \sqrt{ \frac{1-\alpha^2}{g_0} }
 \sqrt{ 1 + \alpha \cos \omega x }
 \tanh\left[
    \frac{\omega}{2} \sqrt{ \frac{1-\alpha^2}{2} } \; X(x)
 \right]
 \end{equation}
 with $X(x)$ being a solution of
  \begin{equation}\label{cuci}
  \tan \left( \frac{ \omega }{ 2 } \sqrt{1-\alpha^2} \; X(x) \right) =
  \sqrt{ \frac{1-\alpha}{1+\alpha} } \; \tan \frac{\omega x}{2}.
\end{equation}
Then, the boundary condition $u_{+}$ is
\begin{equation}\label{boundary2}
u_{+}=\frac{\omega}{2}
\sqrt{ \frac{(1-\alpha^2) (1 + \alpha \cos \omega x)} {g_0}}
 \end{equation}
and $u_{-}=-u_{+}$.

This solution is depicted in the Fig. \ref{darksoliton}.

\subsection{Dark solitons in the photorefractive NLS equation}

There exists many optical materials, like lithium niobate, that exhibits photorefractivity. This nonlinear effect, contrary to the Kerr media, implies saturation, i.e. the function $f$ is increasing and fulfills \cite{Stuart}:
\begin{equation}
f(0)=0,\quad \lim_{u\rightarrow+\infty}f(u)=f_{\infty}<\infty
\end{equation}
The study of photorefractive materials is of high interest, as possess unique features. For instance, they are able to exhibit either self-focusing or self-defocusing in the same crystal; besides, solitons can be generated with low optical power \cite{Kivshar1,PR}.

\begin{figure}[t]
\begin{center}
    \includegraphics[width=12cm]{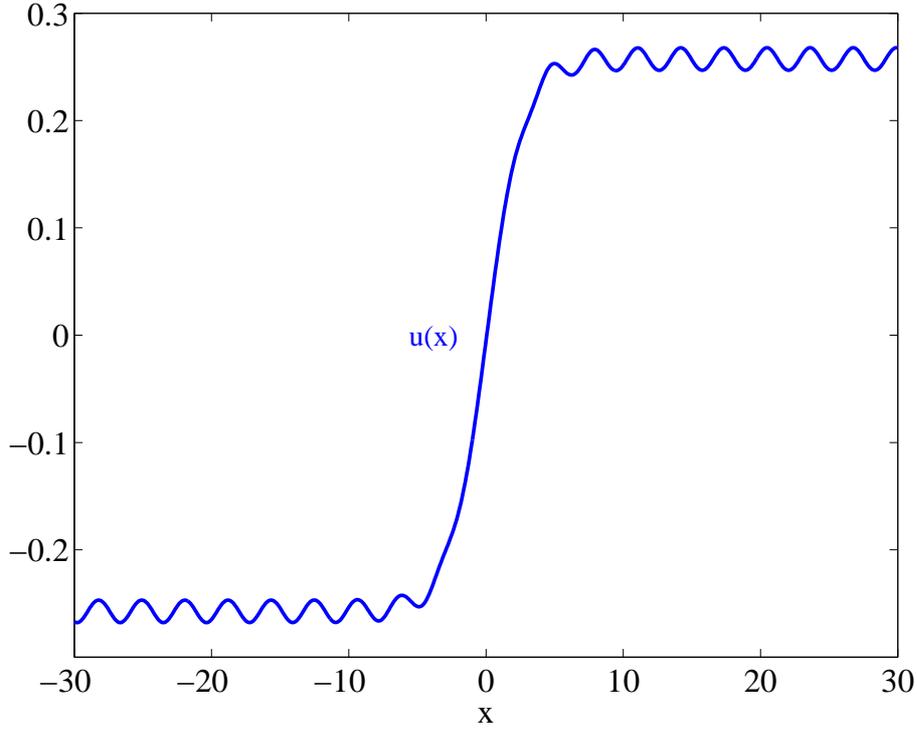}
\caption{[Color Online]  . The dark soliton $u(x)$ calculated for the parameters $\lambda=-0.3$, $g_{\min}=2$, $g_{\max}=8$ of the photorefractive NLS equation. \label{darksoliton2}}
\end{center}
\end{figure}

Photorefractive nonlinearity is given by
\begin{equation}\label{saturation}
f(u)=\frac{u^2}{1+u^2}
\end{equation}
and, consequently, Eq. (\ref{estacionario}) transforms into
\begin{equation}\label{Eq-saturable}
-u_{xx}+\lambda u+g(x)\frac{u^{3}}{1+u^2}=0
\end{equation}
Then, for $f$ given by Eq. (\ref{saturation}), we have one maximum and one minimum for $f'(u)$, given respectively by
\begin{equation}
u=\sqrt{\frac{1}{3}},\quad u=-\sqrt{\frac{1}{3}}
\end{equation}
Thus, we are considering the case $(iii)$ of the previous section. Then, Eq. (\ref{condicion3}) becomes
\begin{equation}\label{cond2}
\lambda-\lambda\frac{g_{\min}}{g_{\max}}+\frac{3\sqrt{3}}{8}g_{\min}\sqrt{\frac{-\lambda}{\lambda+g_{\max}}}>0
\end{equation}
If we expand expression (\ref{cond2}), we obtain the following inequality
\begin{equation}
\lambda^{2}+g_{\max}\lambda+\frac{27}{64}\frac{(g_{\min}g_{\max})^2}{(g_{\min}-g_{\max})^2}>0
\end{equation}
which can be written as
\begin{equation}\label{cond3}
\left(\lambda-\lambda_{1}\right)\left(\lambda-\lambda_{2}\right)>0
\end{equation}
where
\begin{eqnarray}
\lambda_{1}&=&-\frac{g_{\max}}{2}+\frac{g_{\max}}{2}\sqrt{1-\frac{27}{16}\frac{g^2_{\min}}{(g_{\min}-g_{\max})^2}},\\
\lambda_{2}&=&-\frac{g_{\max}}{2}-\frac{g_{\max}}{2}\sqrt{1-\frac{27}{16}\frac{g^2_{\min}}{(g_{\min}-g_{\max})^2}}
\end{eqnarray}
provided that
\begin{equation}\label{ineq1}
g_{\max}\geq \left(1+\frac{\sqrt{27}}{4}\right)g_{\min}
\end{equation}
Thus, from (\ref{cond3}), we obtain that at least for
\begin{equation}\label{sufcond}
\lambda\in (-g_{\max},\lambda_{2}]\cup [\lambda_{1},0)
\end{equation}
Eq.  (\ref{Eq-saturable}) has dark soliton solutions.

Analytical solution of Eq. (\ref{Eq-saturable}) cannot be generally found. A numerical solution for $g(x)=2+6\cos^2(x)$ and $\lambda=-0.3$ is depicted in Fig. \ref{darksoliton2}. For this case, $g_{\min}=2, g_{\max}=8$. We want to note that for these values, the inequality (\ref{ineq1}) is satisfied, being $\lambda_{1}=-0.39$ and $\lambda_{2}=-7.60$. Thus, the sufficient condition (\ref{sufcond}) is satisfied as $\lambda$ belongs to the interval given by (\ref{sufcond}).

In order to find the numerical solution, it has been implemented a fixed point Newton--Raphson algorithm with a finite-differences discretization of the Laplacian, supplemented by anti-periodic boundary conditions.

\section{Conclusions}

In this paper, we have considered the existence of dark solitons or heteroclinic orbits of the nonlinear Schr\"odinger equation with periodic nonlinearity. The proof method relies on some results of the qualitative theory of ordinary differential equations that require some concepts such as upper and lower solutions, topological degree or free homeomorphisms. As an example, we have constructed an analytical black soliton-solution for cubic nonlinearity and a numerical black soliton for photorefractive nonlinearity.

\section*{Acknowledgements}

JBB has been supported by grants: FIS2006-04190 (Ministerio de Educaci\'on y Ciencia, Spain) and PCI08-093 (Consejer\'{\i}a
de Educaci\'on y Ciencia de la Junta de Comunidades de Castilla-La Mancha, Spain). JC acknowledges financial support from the MICINN project FIS2008-04848.

\label{lastpage}

\end{document}